\documentclass[copyright,creativecommons]{eptcs}
\providecommand{\event}{GandALF 2014}

\usepackage{amsmath} 
\usepackage{amssymb} 
\usepackage{amstext}
\usepackage{amsthm}
\usepackage{breakurl} 
\usepackage{color} 
\usepackage{enumerate} 
\usepackage[T1]{fontenc}
\usepackage{ifthen}
\usepackage{multicol}
\usepackage{url}
\usepackage{xspace}

\newcommand{\eg}{\textrm{e.g.,}\@\xspace}

\newcommand{\ie}{\textrm{i.e.,}\@\xspace}

\newcommand{\etal}{\textrm{et al.}\@\xspace}

\newcommand{\COCemph}{\emph{constraint on communication}\@\xspace}
\newcommand{\COC}{constraint on communication\@\xspace}

\newcommand{\PFCemph}{\emph{potential for communication}\@\xspace}
\newcommand{\PFC}{potential for communication\@\xspace}
\newcommand{\PFCD}{potential for direct communication\@\xspace}

\newcommand{\soca}{system of communicating agents\@\xspace}
\newcommand{\socaC}{system~$\C$ of communicating agents\@\xspace}
\newcommand{\socas}{systems of communicating agents\@\xspace}

\newcommand{\lnotation}[4]{
	\def\third:{#3} 
	\def\possiblyone:{} 
	\def\possiblytwo:{~}
	\def\possiblythree:{ }
	\def\divide{\;#1\hspace*{-0pt}( #2\; \mid: \; #4 \, )}
	\def\nodivide{\;#1\hspace*{-0pt}( #2\;\mid\; #3\;:\;#4 \, )}
	\ifx\third\possiblyone\divide
		\else\ifx\third\possiblytwo\divide
		\else \ifx\third\possiblythree\divide
		\else \nodivide\fi\fi\fi}

\newcommand{\biglnotation}[4]{
	\def\third:{#3} 
	\def\possiblyone:{} 
	\def\possiblytwo:{~}
	\def\possiblythree:{ }
	\def\divide{\;#1\hspace*{-0pt}\big( #2\; \mid: \; #4 \, \big)}
	\def\nodivide{\;#1\hspace*{-0pt}\big( #2\;\mid\; #3\;:\;#4 \, \big)}
	\ifx\third\possiblyone\divide
		\else\ifx\third\possiblytwo\divide
		\else \ifx\third\possiblythree\divide
		\else \nodivide\fi\fi\fi}

\newcommand{\bigglnotation}[4]{
	\def\third:{#3} 
	\def\possiblyone:{} 
	\def\possiblytwo:{~}
	\def\possiblythree:{ }
	\def\divide{\;#1\hspace*{-0pt}\bigg( #2\; \mid: \; #4 \, \bigg)}
	\def\nodivide{\;#1\hspace*{-0pt}\bigg( #2\;\mid\; #3\;:\;#4 \, \bigg)}
	\ifx\third\possiblyone\divide
		\else\ifx\third\possiblytwo\divide
		\else \ifx\third\possiblythree\divide
		\else \nodivide\fi\fi\fi}
 
\newcommand{\wplnotation}[4]{
	\def\third:{#3} 
	\def\possiblyone:{} 
	\def\possiblytwo:{~}
	\def\possiblythree:{ }
	\def\divide{\;#1\hspace*{-0pt}\bigg( #2\; \mid: \; #4 \, }
	\def\nodivide{\;#1\hspace*{-0pt}\bigg( #2\;\mid\; #3\;:\;#4 \, }
	\ifx\third\possiblyone\divide
		\else\ifx\third\possiblytwo\divide
		\else \ifx\third\possiblythree\divide
		\else \nodivide\fi\fi\fi}

\newcommand{\grieslnotation}[4]{
	\def\third:{#3} 
	\def\possiblyone:{} 
	\def\possiblytwo:{~}
	\def\possiblythree:{ }
	\def\divide{(#1 #2\; \mid : \; #4 \, )}
	\def\nodivide{(#1 #2\;\mid\; #3\;:\;#4 \, )}
	\ifx\third\possiblyone\divide
		\else\ifx\third\possiblytwo\divide
		\else \ifx\third\possiblythree\divide
		\else \nodivide\fi\fi\fi}

\newcommand{\griesset}[3]{
	\def\second:{#2} 
	\def\possiblyone:{} 
	\def\possiblytwo:{~}
	\def\possiblythree:{ }
	\def\divide{\{#1\; \mid : \; #3 \, \}}
	\def\nodivide{\{#1\;\mid\; #2\;:\;#3 \, \}}
	\ifx\second\possiblyone\domvide
		\else \ifx\second\possiblytwo\divide
		\else \ifx\second\possiblythree\divide
		\else \nodivide\fi\fi\fi}

\newcommand{\set}[1]{\{#1\}}

\newcommand{\sets}[2]{\{#1\; \mid \; #2\}}

\newcommand{\C}{{\cal{C}}}

\newcommand{\monoid}[3]{\big(#1, #2, #3 \big)}

\newcommand{\semiring}[5]{\big(#1, #2, #3, #4, #5\big)}

\newcommand{\Lsemimodule}[3]{\big(_{#1}#2, #3\big)}
\newcommand{\Rsemimodule}[3]{\big(#2_{#1}, #3\big)}

\newcommand{\bigP}[1]{\big( #1 \big)}

\newcommand{\bigA}[1]{\big\langle #1 \big\rangle}

\newlength{\interligne}

\newcommand{\Beginproof}{\dimen123=\linewidth \dimen124=\linewidth
	\advance\dimen123 by -25mm \advance\dimen124 by -5mm
	\advance\dimen123 by -\parindent \advance\dimen124 by -\parindent
	\setlength{\interligne}{\baselineskip}
	\setlength{\baselineskip}{1.2\baselineskip}
    	\begin{tabbing}
    		\hspace*{\parindent}\= \hspace*{5mm}\= \kill \+ \kill}
			\newcommand {\Endproof}
		{\end{tabbing}
    \setlength{\baselineskip}{\interligne}}

\newcommand{\com}[1]{\> \hspace*{15mm}
	$\langle$~\parbox[t]{\dimen123}{ #1 $\rangle$}\\}

\newcommand{\pred}[1]{\>\parbox[t]{\dimen124}{#1}\\}

\newcommand{\hsep}{\quad\&\quad}

\newcommand{\Beginproofitem}{\dimen123=\linewidth \dimen124=\linewidth
	\advance\dimen123 by -20mm \advance\dimen124 by -5mm
	\advance\dimen124 by -\parindent
    	\begin{tabbing}
    		\hspace*{5mm}\= \kill}
			\newcommand {\Endproofitem}
	{\end{tabbing}}

\newcommand {\Begingtabin}{
	\begin{tabbing}
    	\hspace*{\parindent}\= \kill \+ \kill}
		\newcommand {\Endtabin}
	{\end{tabbing}}

\newcommand{\Begingspec}{
	\begin{tabbing}
    	\hspace*{\parindent}\= \hspace*{5mm}\=\hspace*{5mm}\=\hspace*{5mm}\=
			\hspace*{5mm}\=\hspace*{5mm} \kill \+ \kill}
		\newcommand {\Endspec}
	{\end{tabbing}}

\newcommand {\Beginspecitem}{
	\begin{tabbing}
    	\hspace*{5mm}\=\hspace*{5mm}\=\hspace*{5mm}\=
    	\hspace*{5mm}\=\hspace*{5mm} \kill}
		\newcommand {\Endspecitem}
	{\end{tabbing}}

\newcommand{\nln}{@{}l@{}}

\newcommand{\ShowEqSourceStructure}{1}

\newcommand{\eqfrom}[1]{
	\ifthenelse{\ShowEqSourceStructure=1}
	{ \quad {\textcolor{red}{\mbox{#1}}}} {\quad} }

\newcommand{\true}{\textsf{true}}
\newcommand{\false}{\textsf{false}}

\newcommand{\Not}{\neg}

\newcommand{\Or}{\mathrel{\vee}}
\newcommand{\Ors}{\;\Or\;}
\newcommand{\AAnd}{\mathrel{\wedge}}
\newcommand{\nAnd}{\;\AAnd\;}

\newcommand{\mImp}{\;\Longrightarrow\;}  
\newcommand{\mImpl}{\;\Longleftarrow\;}  
\newcommand{\mIff}{\;\Longleftrightarrow\;}

\newcommand{\STbot}{\emptyset}

\newcommand{\STleq}{\subseteq}

\newcommand{\STdiff}{\backslash}

\newcommand{\internalconverse}[1]
	{#1^{\mkern-1mu{}{\raise0.1ex\hbox{\tiny$\smallsmile\,$}}}\kern-0.1em{}}

\newcommand{\RAcomp}{\mathop{\kern-.5pt\raise.3ex\hbox{\footnotesize\rm;}}}

\newcommand{\deq}{\spaces{\stackrel{\mathrm{def}}{=}}}

\newcommand{\spaces}[1]{\,#1\,}

\newcommand{\KA}{Kleene algebra\@\xspace}

\newcommand{\CKA}{concurrent Kleene algebra\@\xspace}
\newcommand{\CKAabbrv}{CKA\@\xspace}

\newcommand{\CKAset}{K}

\newcommand{\CKApar}{*}
\newcommand{\CKAseq}{\raise.3ex\hbox{\,\rm;\,}}
\newcommand{\CKAiterSeqOp}{\text{\scriptsize \textcircled{\raise.3ex\hbox{\,\rm;\,}}}}
\newcommand{\CKAiterParOp}{\text{\scriptsize \textcircled{\raise-.75ex\hbox{\,*\,}}}}
\newcommand{\CKAiterSeq}[1]{{#1}^\CKAiterSeqOp}
\newcommand{\CKAiterPar}[1]{{#1}^\CKAiterParOp}

\newcommand{\cka}{{\cal K}}

\newcommand{\CKAstructure}{\bigP{\CKAset, +, \CKApar, \CKAseq, \CKAiterPar{}, \CKAiterSeq{}, 0, 1}}
\newcommand{\CKAstructurePar}{\bigP{\CKAset, +, \CKApar, \CKAiterPar{}, 0, 1}}
\newcommand{\CKAstructureSeq}{\bigP{\CKAset, +, \CKAseq, \CKAiterSeq{}, 0, 1}}

\newcommand{\CKAle}{\le_{\cka}}

\newcommand{\STIMset}{S}

\newcommand{\STIMplus}{\oplus}
\newcommand{\STIMdot}{\odot}

\newcommand{\Nstim}{\mathfrak{n}}

\newcommand{\Dstim}{\mathfrak{d}}

\newcommand{\stim}{{\cal S}}
\newcommand{\STIMstructure}{\bigP{\STIMset, \STIMplus, \STIMdot, \Dstim, \Nstim}}

\newcommand{\STIMle}{\le_{\stim}}

\newcommand{\rightSemimodule}[1]{right~$#1$-semimodule\@\xspace}
\newcommand{\leftSemimodule}[1]{left~$#1$-semimodule\@\xspace}

\newcommand{\LeftSemimodule}[1]{Left~$#1$-semimodule\@\xspace}

\newcommand{\actOp}{\circ}

\newcommand{\lActSig}{\actOp: \STIMset \times \CKAset \to \CKAset}

\newcommand{\lAct}[2]{{#2} \actOp {#1}}

\newcommand{\outOp}{\lambda}			

\newcommand{\lOutSig}{\outOp: \STIMset \times \CKAset \to \STIMset}	
\newcommand{\lOut}[2]{\outOp(#2,#1)}	

\newcommand{\lOutbig}[2]{\outOp\bigP{#2,#1}}

\newcommand{\orb}[1]{\mathrm{Orb}(#1)}

\newcommand{\orbS}[1]{\mathrm{Orb_{S}}(#1)}

\newcommand{\CCKA}{Communicating Concurrent Kleene Algebra\@\xspace}
\newcommand{\CCKAabbrv}{C$^2$KA\@\xspace}

\newcommand{\CCKAstructure}{\bigP{\stim, \cka}}

\newcommand{\ActSemimodule}{\Lsemimodule{\stim}{\CKAset}{+}}
\newcommand{\OutSemimodule}{\Rsemimodule{\cka}{\STIMset}{\STIMplus}}

\newcommand{\STIMbasic}{\STIMset_{b}} 

\newcommand{\Agent}[1]{\mathsf{#1}}

\newcommand{\agent}[2]{\Agent{#1} = \bigA{#2}}

\newcommand{\comm}[2]{\mathrel{{\to}_{#1}^{#2}}}

\newcommand{\STIMcommD}[2]{#1 \comm{\stim}{} #2}
\newcommand{\STIMcommN}[3]{#1 \comm{\stim}{#3} #2}
\newcommand{\STIMcomm}[2]{\STIMcommN{#1}{#2}{*}}
\newcommand{\notSTIMcomm}[2]{\Not(\STIMcomm{#1}{#2})}

\newcommand{\env}{{\cal{E}}}
\newcommand{\ENVcommD}[2]{#1 \comm{\env}{} #2}
\newcommand{\ENVcommN}[3]{#1 \comm{\env}{#3} #2}
\newcommand{\ENVcomm}[2]{\ENVcommN{#1}{#2}{*}}

\newcommand{\pfcD}[2]{#1 \leadsto #2}

\newcommand{\pfc}[2]{#1  \leadsto^{*} #2}
\newcommand{\notpfc}[2]{\Not(\pfc{#1}{#2})}

\newcommand{\depOp}{\mathrm{R}}
\newcommand{\depOpTC}{\depOp^{+}}
\newcommand{\dep}[2]{#2 \,\depOp\, #1}
\newcommand{\depTC}[2]{#2 \,\depOpTC\, #1}
\newcommand{\indep}[2]{\Not(#2 \,\depOp\, #1)}

\newtheorem{definition}{Definition}
\newtheorem{proposition}{Proposition}

\newcommand{\keywords}[1]{\par\addvspace\baselineskip\noindent
							\textbf{Keywords:}\enspace\ignorespaces#1}

\begin{document}

\title{A Formulation of the Potential for \\ Communication Condition using \CCKAabbrv\footnote{This research is supported by the Natural Sciences and Engineering Research Council of Canada (NSERC) through the grant~RGPIN~2014-06115 and the NSERC~PGS~D program.}}
\author{Jason Jaskolka \qquad\qquad Ridha Khedri
\institute{Department of Computing and Software,	
			Faculty of Engineering 					\\
		  	McMaster University, 					
		  	Hamilton, Ontario, Canada	}
\email{\quad jaskolj@mcmaster.ca \qquad\qquad khedri@mcmaster.ca}}
\def\titlerunning{A Formulation of the Potential for Communication Condition using \emph{\CCKAabbrv}}
\def\authorrunning{J. Jaskolka and R. Khedri}
		
\maketitle

\begin{abstract}
\vspace{-0.1in}
	An integral part of safeguarding \socas from covert channel communication is having the ability to identify when a covert channel may exist in a given system and which agents are more prone to covert channels than others. In this paper, we propose a formulation of one of the necessary conditions for the existence of covert channels: the \PFC condition. Then, we discuss when the \PFC is preserved after the modification of system agents in a potential communication path. Our approach is based on the mathematical framework of \CCKA~(\CCKAabbrv). While existing approaches only consider the potential for communication via shared environments, the approach proposed in this paper also considers the potential for communication via external stimuli. 
	
\end{abstract}

\vspace{-0.1in}	
\section{Introduction and Motivation} 
\label{sec:introduction}
\vspace{-0.1in}
Today we are faced with large and complex networks, consisting of numerous communicating agents which have the ability to harbour countless covert communication channels. A covert channel refers to any communication means that allows an agent to transfer information in a manner that violates a system's security policy~\cite{DoD1985aa}. We can imagine a complex network of communicating agents organised in such a way that covert communication can be widespread across the entire network and which can utilise a number of different communication mediums, channels, and techniques as depicted by the perception of covert channel communication given in~\cite{Jaskolka2011aa}. The existence of covert communication channels introduces a number of security concerns such as confidentiality concerns and economical concerns. In~\cite{Jaskolka2012aa}, we presented a set of informal conditions which are necessary for the existence of covert communication channels in \socas. In such systems, if there exists a covert communication channel, then the \COCemph and \PFCemph conditions are satisfied. In this paper, we focus on providing a formulation of the \PFC condition. The \PFC condition states that if there exists the possibility for information to flow from one agent to another through the synchronisation and sequencing of events in a system, then the agents have the potential for communication. 

Currently, covert channels are poorly understood~\cite{Jaskolka2011aa}. There are shortcomings in the science, mathematics, and fundamental theory to deal with covert channels in modern computer systems~\cite{DoHS2009aa}. One of the first steps towards uncovering whether covert channels can exist in a given \soca is to identify which agents have the \PFC. There are a limited number of existing approaches for identifying \PFC in \socas. Those that do exist are typically information theoretic approaches (\eg~\cite{Gianvecchio2007aa,Giles2002aa,Giles1999aa,Helouet2010aa,Lowe2002aa,Martin2006aa,Millen1987aa,Millen1989aa,Moskowitz1994aa}). These approaches attempt to identify \PFC by looking for positive capacity channels that may exist among system agents. However, the notion of channel capacity is an insufficient stand-alone measure for the existence of covert channels~\cite{Moskowitz1994aa}. As motivation for this argument, an example of a zero capacity channel is given in~\cite{Moskowitz1994aa}, on which any message can be sent, thus illustrating that knowing that the capacity is zero does not ensure that there is no \PFC. Other existing approaches view \PFC from the perspective of information flows (\eg~\cite{Kemmerer1983aa,Kemmerer1991aa}). However, these approaches only consider communication via shared environments by examining the dependencies between shared events. 

The formulation proposed in this paper is based on the mathematical framework of \CCKA~(\CCKAabbrv)~\cite{Jaskolka2013aa,Jaskolka2014aa} which is an extension of the work of Hoare \etal~\cite{Hoare2009aa,Hoare2009ab,Hoare2010aa,Hoare2011aa}. This framework provides a means for specifying \socas and allows for the separation of communicating and concurrent behaviour in a system and its environment. Because of this, we are able to consider the potential for communication amongst agents from two complementary perspectives. First, we consider the \PFC via external stimuli which examines how stimuli generated from one agent in the system are able to influence the behaviour of other agents in the system. Second, we consider the \PFC via shared environments which studies how communication can occur through shared events/variables and the dependencies between them. By formulating the \PFC condition for covert channel existence using \CCKAabbrv, we can formally verify the satisfaction of the condition for a given \soca. The proposed formulation can serve as the basis for developing effective and efficient mechanisms for mitigating covert channels in \socas. This can allow us to strengthen the design of systems so that they are more robust against covert channels. 

The remainder of this paper is organised as follows. Section~\ref{sec:background} gives the required background of covert channel communication and \CCKAabbrv. Section~\ref{sec:formulating_the_potential_for_communication_condition} provides a formulation of the \PFC condition using \CCKAabbrv. Section~\ref{sec:discussion} discusses the proposed formulation along with related work. Finally, Section~\ref{sec:conclusion_and_future_work} draws conclusions and provides the highlights of our future work.

\vspace{-0.1in}
\section{Background} 
\label{sec:background}
\vspace{-0.1in}
\subsection{Covert Channel Communication}
\label{sub:covert_channel_communication}

A covert channel is any communication means that allows information to be transferred by system agents in a manner that violates the system's security policy~\cite{DoD1985aa}. Typically, covert channels are hidden from the view of third party observers. In this way, the use of covert channels often results in third-party observers not even necessarily being aware that any communication is taking place at all.

Today, systems comprise of broad and heterogeneous communication networks with many interacting agents. This yields numerous possibilities for covert channels. Systems consist of physical networks, virtual networks, and even social networks and can be spread across a variety of application domains, each with their own security concerns with varying implications and priorities. Because of the scale and complexity of such systems, the need for a systematic analysis of \socas for the existence of covert channels is becoming increasingly important. 

Covert channels can be classified as either \emph{protocol-based}, \emph{environment-based}, or both~\cite{Jaskolka2010aa}. A protocol-based covert channel is a communication means that uses a communication protocol to convey messages that violate a security policy whereas an environment-based covert channel is a communication means that uses environmental resources, functionalities, or features, including timing information, to convey messages that violate a security policy.


\subsection{\CCKA}
\label{sub:c2ka}

\CCKA~(\CCKAabbrv) extends the algebraic foundation of Concurrent Kleene Algebra~(\CKAabbrv), proposed by Hoare \etal~\cite{Hoare2009aa,Hoare2009ab,Hoare2010aa,Hoare2011aa}, with the notions of semimodules and stimulus structures to capture the influence of external stimuli on the behaviour of system agents. For a full account of \CCKAabbrv, the reader is referred to~\cite{Jaskolka2013aa,Jaskolka2014aa}. 

A \emph{monoid} is a mathematical structure~$\monoid{S}{\cdot}{1}$ consisting of a nonempty set~$S$, together with an associative binary operation~$\cdot$ and a distinguished constant~$1$ which is the identity with respect to~$\cdot$. A monoid is called \emph{commutative} if~$\cdot$ is commutative and a monoid is called \emph{idempotent} if~$\cdot$ is idempotent.

A \emph{semiring} is a mathematical structure~$\semiring{S}{+}{\cdot}{0}{1}$ consisting of a commutative monoid~$\monoid{S}{+}{0}$  and a monoid~$\monoid{S}{\cdot}{1}$ such that operator~$\cdot$ distributes over operator~$+$. We say that element~$0$ is \emph{multiplicatively absorbing} if it annihilates~$S$ with respect to~$\cdot$. We say that a semiring is \emph{idempotent} if operator~$+$ is idempotent. Every idempotent semiring has a natural partial order~$\le$ on~$S$ defined by~$a \le b \!\mIff\! a + b = b$. Operators~$+$ and~$\cdot$ are isotone on both the left and the right with respect to~$\le$.

A \emph{\KA} is mathematical structure that extends the notion of idempotent semirings with the addition of a unary operator for finite iteration. Kleene algebras are most commonly known for generalising the operations of regular expressions.

\begin{definition}[\LeftSemimodule{\stim} -- \eg~\cite{Hebisch1993aa}]
\label{def:semimodule}
	Let~$\stim = \semiring{\STIMset}{+}{\cdot}{0_\stim}{1}$ be a semiring and~$\cka = \monoid{\CKAset}{\STIMplus}{0_\cka}$ be a commutative monoid. We call~$\Lsemimodule{\stim}{\CKAset}{\STIMplus}$ a \emph{\leftSemimodule{\stim}} if there exists a mapping~$\STIMset \times \CKAset \to \CKAset$ denoted by juxtaposition such that for all~$s,t \in \STIMset$ and~$a,b \in \CKAset$
	\begin{enumerate}[(i)]
		\begin{minipage}[t]{0.4\linewidth}   
			\item \label{def:SM_dist_Kplus}
				$s(a \STIMplus b) = sa \STIMplus sb$
			\item \label{def:SM_dist_Splus}
				$(s + t)a = sa \STIMplus sb$
			\item \label{def:SM_assoc_seq}
				$(s \cdot t)a = s(ta)$
		\end{minipage}
	  	\begin{minipage}[t]{0.6\linewidth}
			\item \label{def:SM_id}
				$\Lsemimodule{\stim}{\CKAset}{\STIMplus}$ is called \emph{unitary} if it also satisfies~$1a = a$
			\item \label{def:SM_zero}
				$\Lsemimodule{\stim}{\CKAset}{\STIMplus}$ is \emph{zero-preserving} if it also satisfies~$0_{\stim}a = 0_\cka$
			\textcolor{white}{\item}      
	  	\end{minipage}
	\end{enumerate}
\end{definition}

A \rightSemimodule{\stim} can be defined analogously. 

Concurrent Kleene algebra is an algebraic framework that extends \KA by offering operators for sequential and concurrent composition, along with those for choice and finite iteration. 

\begin{definition}[Concurrent Kleene Algebra -- \eg~\cite{Hoare2009aa}]
\label{def:CKA}
	A \emph{\CKA (\CKAabbrv)} is a structure~$\cka \deq \CKAstructure$ where~$\CKAstructurePar$ and~$\CKAstructureSeq$ are Kleene algebras linked by the \emph{exchange axiom} given by~$(a \CKApar b) \CKAseq (c \CKApar d) \CKAle (b \CKAseq c) \CKApar (a \CKAseq d)$.
\end{definition}

Within the context of agent behaviours,~$\CKAset$ represents a set of possible agent behaviours. The operator~$+$ is interpreted as a choice between two behaviours, the operator~$\CKApar$ is interpreted as a parallel composition of two behaviours, and the operator~$\CKAseq$ is interpreted as a sequential composition of two behaviours. The operators~$\CKAiterPar{}$ and~$\CKAiterSeq{}$ are interpreted as finite parallel iteration and finite sequential iteration, respectively. The element~$0$ represents the behaviour of the \emph{inactive agent} and the element~$1$ represents the behaviour of the \emph{idle agent} just as in many process calculi. Moreover, an agent behaviour~$a$ is a \emph{sub-behaviour} of an agent behaviour~$b$, denoted~$a \CKAle b$, if and only if~$a + b = b$. In this way, the exchange axiom intuitively expresses a divide-and-conquer mechanism for how parallel composition may be sequentially implemented on a machine.

When we speak of agents and agent behaviours, we write~$\agent{A}{a}$ where~$\Agent{A}$ is the name given to the agent and~$a \in \CKAset$ is the agent behaviour. For~$\agent{A}{a}$ and~$\agent{B}{b}$, we write~$\Agent{A+B}$ to denote the agent~$\bigA{a+b}$. In a sense, we extend the operators on behaviours of~$\CKAset$ to their corresponding agents. In this way, an agent is defined by simply describing its behaviour. Because of this, we may use the terms agents and behaviours interchangeably.

\begin{definition}[Stimulus Structure -- \eg~\cite{Jaskolka2014aa}]
	\label{def:stimulus_structure}
	Let~$\stim \deq \STIMstructure$ be an idempotent semiring with a multiplicatively absorbing~$\Dstim$ and identity~$\Nstim$. We call~$\stim$ a \emph{stimulus structure}.
\end{definition}

Within the context of external stimuli,~$\STIMset$ is the set of stimuli which may be introduced to a system. A stimulus can be thought of as an event that has the potential to affect agent behaviour. The operator~$\STIMplus$ is interpreted as a choice between two stimuli and the operator~$\STIMdot$ is interpreted as a sequential composition of two stimuli. The element~$\Dstim$ represents the \emph{deactivation stimulus} which influences all agents to become inactive and the element~$\Nstim$ represents the \emph{neutral stimulus} which has no influence on the behaviour of all agents. We say that~$s \in \STIMset$ is a \emph{basic stimulus} if it is indivisible with regard to the~$\STIMdot$ operator (\ie~$\biglnotation{\forall}{t}{}{(t|s) \mImp (t = \Nstim \Ors t = s)}$ and~$\biglnotation{\forall}{t,r}{}{(s|(t \STIMdot r)) \mImp (s|t \Ors s|r)}$ where the division operator~$|$ is defined by~$x|y \mIff \lnotation{\exists}{z}{}{y = x \STIMdot z}$). We denote the set of all basic stimuli as~$\STIMbasic$. Furthermore, a stimulus~$s$ is a \emph{sub-stimulus} of a stimulus~$t$, denoted~$s \STIMle t$, if and only if~$s \STIMplus t = t$.

\begin{definition}[\CCKA\ -- \eg~\cite{Jaskolka2014aa}]
	\label{def:C2KA}
	A \emph{\CCKA (\CCKAabbrv)} is a system~$\CCKAstructure$, where $\stim = \STIMstructure$ is a stimulus structure and~$\cka = \CKAstructure$ is a \emph{\CKAabbrv} such that~$\ActSemimodule$ is a unitary and zero-preserving \emph{\leftSemimodule{\stim}} with mapping~$\lActSig$ and~$\OutSemimodule$ is a unitary and zero-preserving \emph{\rightSemimodule{\cka}} with mapping~$\lOutSig$, and where the following axioms are satisfied for all~$a,b,c \in \CKAset$ and~$s,t \in \STIMset$:
	\begin{multicols}{2}
	\begin{enumerate}[(i)]
		\item \label{def:cascading_axiom}
			$\lAct{(a \CKAseq b)}{s} = (\lAct{a}{s}) \CKAseq \bigP{\lAct{b}{\lOut{a}{s}}}$
		\item \label{def:cascading_output_axiom}
			$c \CKAle a \Ors (\lAct{a}{s}) \CKAseq \bigP{\lAct{b}{\lOut{c}{s}}} = 0$
		\item \label{def:sequential_output_axiom}
			$\lOut{a}{s \STIMdot t} =  \lOutbig{(\lAct{a}{t})}{s} \STIMdot \lOut{a}{t}$
	\end{enumerate}
	\end{multicols}
\end{definition}

A \CCKAabbrv consists of two semimodules which describe how the stimulus structure~$\stim$ and the \CKAabbrv~$\cka$ mutually act upon one another. In this way, the response invoked by a stimulus on the behaviour of an agent is characterised as a next behaviour and a next stimulus. The \leftSemimodule{\stim}~$\ActSemimodule$ describes how the stimulus structure~$\stim$ acts upon the \CKAabbrv~$\cka$ via the \emph{next behaviour mapping}~$\actOp$ and the \rightSemimodule{\cka}~$\OutSemimodule$ describes how the \CKAabbrv~$\cka$ acts upon the stimulus structure~$\stim$ via the \emph{next stimulus mapping}~$\outOp$. Axiom~(\ref{def:cascading_axiom}) describes the interaction of the next behaviour mapping~$\actOp$ with the sequential composition operator~$\CKAseq$ for agent behaviours. Axiom~(\ref{def:cascading_output_axiom}) states that when an external stimulus is introduced to the sequential composition~$(a \CKAseq b)$, then the stimulus cascaded to~$b$ must be generated by a sub-behaviour of~$a$. In this way, Axiom~(\ref{def:cascading_output_axiom}) ensures consistency between the next behaviour and next stimulus mappings with respect to the sequential composition of agent behaviours. Finally, Axiom~(\ref{def:sequential_output_axiom}) describes the interaction of the next stimulus mapping~$\outOp$ with the sequential composition operator~$\STIMdot$ for external stimuli. This can be viewed as the analog of Axiom~(\ref{def:cascading_axiom}) with respect to the next stimulus mapping~$\outOp$ when considering the action of~$\OutSemimodule$. When examining the effects of external stimuli on agent behaviours, it is important to note that every stimulus \emph{invokes a response} from an agent. When the behaviour of an agent changes as a result of the response, we say that the stimulus \emph{influences} the behaviour of the agent. Moreover, we say that a \CCKAabbrv is \emph{without reactivation} if~$\lnotation{\forall}{s}{s \in \STIMset \STdiff \set{\Dstim}}{\lAct{1}{s} = 1}$.

We recall the notions of orbits, strong orbits, and fixed points from the mathematical theory of monoids acting on sets~\cite{Kilp2000aa}. Let~$\ActSemimodule$ be the unitary and zero-preserving \leftSemimodule{\stim} of a \CCKAabbrv and let~$a \in \CKAset$. The \emph{orbit} of~$a$ in~$\stim$ is the set~$\orb{a} = \sets{\lAct{a}{s}}{s \in \STIMset}$ and represents the set of all possible behavioural responses from an agent behaving as~$a$ to any stimulus from~$\stim$.
The \emph{strong orbit} of~$a$ in~$\stim$ is the set~$\orbS{a} = \sets{b \in \CKAset}{\orb{b} = \orb{a}}$. Two agents are in the same strong orbit if and only if their orbits are identical. This is to say, if an agent behaving as~$a$ is influenced by a stimulus to behave as~$b$, then there exists a stimulus which influences the agent, now behaving as~$b$, to revert back to its original behaviour~$a$. Furthermore, if~$a$ and~$b$ are in the same strong orbit, then~$\lnotation{\exists}{s,t}{s,t \in \STIMset}{\lAct{a}{s} = b \nAnd \lAct{b}{t} = a}$. Lastly, we say that the element~$a \in \CKAset$ is a \emph{fixed point behaviour} if~$\lnotation{\forall}{s}{s \in \STIMset \STdiff \set{\Dstim}}{\lAct{a}{s} = a}$. In other words,~$a$ is a fixed point behaviour if it is not influenced by any external stimuli other than the deactivation stimulus~$\Dstim$.



\section{Formulating the Potential for Communication Condition}
\label{sec:formulating_the_potential_for_communication_condition}
The \PFCemph condition is introduced as one of the two necessary conditions for covert channel existence in~\cite{Jaskolka2012aa}. 
The condition reads:
\begin{quote}
	If there exists an agent acting as a source of information and an agent acting as an information sink, such that the source and sink agents are different, and if there exists a pattern of communication allowing for information to transfer from the source to the sink through the synchronisation and sequencing of events, then the source and sink agents have a \PFC.
\end{quote}
In this section, we propose a formulation of the \PFC condition using \CCKAabbrv. In what follows, we adopt the notion of communication used in~\cite{Milner1989aa}, where each interaction (direct or indirect) of an agent with its neighbouring agents is called a \emph{communication}. We examine the \PFC from two complementary perspectives, namely the external stimuli perspective and the shared environment perspective, consistent with the view of communication introduced in~\cite{Jaskolka2013aa,Jaskolka2014aa}. Throughout the following subsections, let~$\C$ be a collection of agents. We call~$\C$ a \emph{\soca}.

\subsection{Formulating Potential for Communication via External Stimuli}
\label{sub:potential_for_communication_via_external_stimuli}

When considering communication in a \soca from the perspective of external stimuli, we need to look at the interactions of the agents. In a given \soca, each agent is subjected to each external stimulus. This means that when an agent generates a stimulus, it is broadcasted to all other agents and a response is invoked. However, it is not the case that the behaviour of each agent will be influenced by the stimulus. Only when a stimulus that is generated by an agent influences  (\ie does not fix) the behaviour of another agent do we say that \emph{communication via external stimuli} has taken place. 

Let~$\Agent{A}, \Agent{B} \in \C$ such that~$\Agent{A} \neq \Agent{B}$. We say that~$\agent{A}{a}$ has the \emph{\PFCD via external stimuli} with~$\agent{B}{b}$ (denoted by~$\STIMcommD{\Agent{A}}{\Agent{B}}$) if and only if~$\biglnotation{\exists}{s,t}{s,t \in \STIMbasic \nAnd t \STIMle \lOut{a}{s} }{\lAct{b}{t} \neq b}$ where~$\STIMbasic$ is the set of all basic stimuli. This means that if there exists a basic sub-stimulus that is generated by~$\Agent{A}$ that influences the behaviour of~$\Agent{B}$, then there is a \PFCD via external stimuli from~$\Agent{A}$ to~$\Agent{B}$. We say that~$\Agent{A}$ has the \emph{\PFC via external stimuli with~$\Agent{B}$ using at most~$n$ basic stimuli} (denoted by~$\STIMcommN{\Agent{A}}{\Agent{B}}{n}$) if and only if~$\biglnotation{\exists}{\Agent{C}}{\Agent{C} \in \C \nAnd \Agent{C} \neq \Agent{A} \nAnd \Agent{C} \neq \Agent{B}}{\STIMcommN{\Agent{A}}{\Agent{C}}{(n-1)} \nAnd \STIMcommD{\Agent{C}}{\Agent{B}}}$. More generally, we say that~$\Agent{A}$ has the \emph{\PFC via external stimuli} with~$\Agent{B}$ (denoted by~$\STIMcomm{\Agent{A}}{\Agent{B}}$) if and only if~$\biglnotation{\exists}{n}{n \ge 1}{\STIMcommN{\Agent{A}}{\Agent{B}}{n}}$. This means that when~$\STIMcomm{\Agent{A}}{\Agent{B}}$, there is a sequence of external stimuli of arbitrary length which allows for information to be transferred from~$\Agent{A}$ to~$\Agent{B}$ in the \socaC.

We say that two subsets~$X_1$ and~$X_2$ of~$\C$ form a partition of~$\C$ if and only if~$X_1 \cap X_2 \!=\! \STbot$ and~$X_1 \cup X_2 \!=\! \C$. A \socaC is said to be \emph{stimuli-connected} if and only if for every~$X_1$ and~$X_2$ that form a partition of~$\C$, we have~$\lnotation{\exists}{\Agent{A},\Agent{B}}{\Agent{A} \in X_1 \nAnd \Agent{B} \in X_2}{\STIMcomm{\Agent{A}}{\Agent{B}} \Ors \STIMcomm{\Agent{B}}{\Agent{A}}}$. Otherwise, we say that~$\C$ is \emph{stimuli-disconnected}. This means that in a stimuli-connected system, every agent is a participant, either as the source or sink, of at least one direct communication via external stimuli.

We say that an agent~$\Agent{A} \in \C$ is a \emph{communication fixed point} if and only if~$\biglnotation{\forall}{\Agent{B}}{\Agent{B} \in \C \STdiff \set{\Agent{A}}}{\notSTIMcomm{\Agent{A}}{\Agent{B}}}$. Obviously, a communication fixed point does not have the \PFC via external stimuli with any other agent. Thus, it is plain to see that an agent~$\agent{A}{0}$ is a communication fixed point since for all~$s \in \STIMset$ we have~$\lOut{0}{s} = \Dstim$ and since~$\Dstim$ is not a basic stimulus, it cannot have the \PFC via external stimuli with any other agent. Additionally, if~$\STIMcomm{\Agent{A}}{\Agent{B}}$, then the potential communication path from~$\Agent{A}$ to~$\Agent{B}$ contains at most one communication fixed point that is~$\Agent{B}$. 

An agent~$\Agent{A} \in \C$ is said to be \emph{universally influential} if and only if~$\biglnotation{\forall}{\Agent{B}}{\Agent{B} \in \C \STdiff \set{\Agent{A}}}{\STIMcomm{\Agent{A}}{\Agent{B}}}$. Every stimulus that is generated by a universally influential agent influences the behaviour, either directly or indirectly, of each other agent in the system. In this way, a universally influential agent is the dual of a communication fixed point and therefore it is obvious that a communication fixed point cannot be universally influential. 

\begin{proposition}
\label{prop:universally_influential_connected}
	A \soca that contains a universally influential agent is stimuli-connected.
	
	\begin{proof}
		Assume~$\C$ is a stimuli-disconnected system and let~$\Agent{C} \in \C$ be universally influential. Then, using the definition of a stimuli-disconnected system, instantiation with~$\Agent{B} = \Agent{C}$, and the definition of universally influential, we have that either~$\C$ is stimuli-connected or~$\Agent{C}$ is not universally influential which is a contradiction to the assumption that~$\C$ is stimuli-disconnected and~$\Agent{C}$ is universally influential. The detailed proof can be found in Appendix~\ref{sub:detailed_proof_of_proposition_universally_influential_connected}.
	\end{proof}
\end{proposition}

\begin{proposition}
\label{prop:fixed_point}
	Let~$\agent{A}{a}$ be an agent such that~$a$ is a fixed point behaviour. Then, there does not exist an agent~$\Agent{B}$ that has the \PFC via external stimuli with~$\Agent{A}$.
	
	\begin{proof}
		The proof is straightforward using the definition of~$\STIMcommD{}{}$.
	\end{proof}
\end{proposition}

In Proposition~\ref{prop:fixed_point}, we have that no agent has the \PFC via external stimuli with an agent that has a fixed point behaviour. This is due to the fact that if an agent has a fixed point behaviour, then it is not influenced by any external stimuli and therefore communication with that agent via external stimuli is not possible.

\begin{proposition}
\label{prop:stim_comm_plus}	
	Let~$\agent{A}{a}$,~$\agent{B}{b}$, and~$\agent{C}{c}$ be agents in~$\C$.
	\begin{enumerate}[(i)]
		\item \label{prop:stim_comm_plus_sink}
			If~$\STIMcommD{\Agent{B}}{\Agent{C}}$ then~$\STIMcommD{(\Agent{A + B})}{\Agent{C}}$.
		\item \label{prop:stim_comm_plus_source}
			If~$\STIMcommD{\Agent{A}}{\Agent{B}}$ then~$\STIMcommD{\Agent{A}}{(\Agent{B + C})}$ only if~$\lnotation{\forall}{t}{t \in \STIMbasic}{\Not(\lAct{c}{t} \CKAle b + c)}$.
	\end{enumerate}

	\begin{proof}
		 The proof of~(\ref{prop:stim_comm_plus_sink}) uses the definition of~$\STIMcommD{}{}$, the distributivity of~$\outOp$ over~$+$, the definition of~$\STIMle$, and isotony of~$=$. The proof of~(\ref{prop:stim_comm_plus_source}) involves the definition of~$\STIMcommD{}{}$ and the distributivity of~$\actOp$ over~$+$, weakening, the definition of~$\CKAle$, and isotony of~$=$. The detailed proofs can be found in Appendix~\ref{sub:detailed_proof_of_proposition_stim_comm_plus}. 
	\end{proof}
\end{proposition}

Proposition~\ref{prop:stim_comm_plus} shows how the \PFC via external stimuli can be preserved when we introduce non-determinism among agents. Specifically, Proposition~\ref{prop:stim_comm_plus}(\ref{prop:stim_comm_plus_sink}) states that when non-determinism is added at the source of a potential communication path via external stimuli, the \PFC via external stimuli is always preserved. Intuitively, this is the case since there can always be a sub-stimulus generated by the source which results from~$\Agent{B}$ that can preserve the \PFC via external stimuli with~$\Agent{C}$. On the other hand, Proposition~\ref{prop:stim_comm_plus}(\ref{prop:stim_comm_plus_source}) states that when non-determinism is added at the sink of a potential communication path via external stimuli, the \PFC is preserved only if there does not exist any basic stimulus that influences~$\Agent{C}$ to behave as a sub-behaviour of~$\Agent{B + C}$. This condition ensures that~$\Agent{B + C}$ cannot have a fixed point behaviour. If the non-determinism that is introduced causes a fixed point behaviour, then there will no longer be any \PFC as stated by Proposition~\ref{prop:fixed_point}.


\subsection{Formulating Potential for Communication via Shared Environments}
\label{sub:potential_for_communication_via_shared_environments}

The examination of communication via shared environments, either through shared variables, resources, or functionalities, has been the topic of study for a number of existing techniques for covert channel and information flow analysis (\eg~\cite{Kemmerer1983aa,Kemmerer1991aa,Sabri2009aa,Shieh1999aa,Wang2005aa}). When formulating the \PFC via shared environments, we are interested in finding if a particular agent has the ability to alter an element of the environment that it shares with a neighbouring agent such that the neighbouring agent is able to observe the alteration that was made.

Since the proposed formulation is based on \CCKAabbrv which is an extension of \CKAabbrv, we utilise the mechanisms provided by \CKAabbrv to formulate the \PFC via shared environments. Similar to what is done with existing information flow techniques for formulating the \PFC via shared environments, we study the dependencies between events that are shared amongst system agents.

In what follows, let~$\bigP{\CKAset, +}$ be an aggregation algebra~\cite{Hoare2009ab,Hoare2010aa,Hoare2011aa} where~$\CKAset$ is a set of agent behaviours and~$+$ is the choice between agent behaviours and let~$a,b,c \in \CKAset$. A \emph{dependence relation} on~$\bigP{\CKAset, +}$ is a bilinear relation~$\depOp \STleq \CKAset \times \CKAset$ (\ie~$\dep{c}{(a + b)} \mIff (\dep{c}{a} \Ors \dep{c}{b})$ and~$\dep{(b + c)}{a} \mIff (\dep{b}{a} \Ors \dep{c}{a})$) where~$\dep{b}{a}$ denotes that the behaviour~$b$ depends on the behaviour~$a$. Such a dependence relation may be a definition-reference relation between program variables in the specifications of agent behaviours. We additionally assume that~$\indep{0}{a}$ and~$\indep{a}{0}$ and~$\indep{1}{a}$ and~$\indep{a}{1}$ for every~$a \in \CKAset$. These are rather natural assumptions since the inactive and idle behaviours depend on nothing and nothing depends on them. Such assumptions are additionally made by Hoare \etal~\cite{Hoare2011aa}. For the purpose of this formulation, we assume that such a dependence relation~$\depOp$ is given.

For~$\Agent{A},\Agent{B} \in \C$ such that~$\Agent{A} \neq \Agent{B}$, we say~$\agent{A}{a}$ has the \emph{\PFCD via shared environments} with~$\agent{B}{b}$ (denoted by~$\ENVcommD{\Agent{A}}{\Agent{B}}$) if and only if~$\dep{b}{a}$. Furthermore, we say that~$\Agent{A}$ has the \emph{\PFC via shared environments} with~$\Agent{B}$ (denoted by~$\ENVcomm{\Agent{A}}{\Agent{B}}$) if and only if~$\depTC{b}{a}$ where~$\depOpTC$ is the transitive closure of the given dependence relation. This means that if two agents respect the given dependence relation, then there is a \PFC via shared environments. 

\begin{proposition}
\label{prop:env_comm_plus}	
	Let~$\C$ be a \soca and let~$\Agent{A}, \Agent{B}, \Agent{C} \in \C$.
	\begin{multicols}{2}
	\begin{enumerate}[(i)]
		\item \label{prop:env_comm_plus_sink}
			If~$\ENVcommD{\Agent{B}}{\Agent{C}}$ then~$\ENVcommD{(\Agent{A} + \Agent{B})}{\Agent{C}}$.
		\item \label{prop:env_comm_plus_source}
			If~$\ENVcommD{\Agent{A}}{\Agent{B}}$ then~$\ENVcommD{\Agent{A}}{(\Agent{B} + \Agent{C})}$.
	\end{enumerate}
	\end{multicols}

	\begin{proof}
		The proofs are straightforward from the definition of~$\ENVcommD{}{}$ and the bilinearity of the dependence relation~$\depOp$.
	\end{proof}
\end{proposition}

Proposition~\ref{prop:env_comm_plus} shows that the \PFC via shared environments is preserved when we introduce non-determinism at the source or the sink of a potential communication path via shared environments. If we know that there exists a dependency between two agent behaviours~$a$ and~$b$, then given a choice between~$b$ and any other behaviours, it is possible to choose to behave as~$b$ in order to preserve the dependency. While this is not always the case, it is important to note that we are focussed on identifying the \PFC, which means that if it is possible for an agent to choose a behaviour which yields the \PFC, then in general the \PFC exists.


\vspace{-0.1in}
\subsection{A Formulation of the Potential for Communication Condition}
\label{sub:a_formulation_of_the_potential_for_communication_condition}

By combining the definitions of \PFC via external stimuli and via shared environments, we obtain a formulation of the \PFC condition for covert channel existence.

For~$\Agent{A},\Agent{B} \in \C$, we say that~$\Agent{A}$ has the \emph{\PFCD} with~$\Agent{B}$ (denoted by~$\pfcD{\Agent{A}}{\Agent{B}}$) if and only if~$\STIMcommD{\Agent{A}}{\Agent{B}} \Ors \ENVcommD{\Agent{A}}{\Agent{B}}$. We say that~$\Agent{A}$ has the \emph{\PFC} with~$\Agent{B}$ (denoted by~$\pfc{\Agent{A}}{\Agent{B}}$) if and only if~$\pfcD{\Agent{A}}{\Agent{B}} \Ors \biglnotation{\exists}{\Agent{C}}{\Agent{C} \in \C}{\pfcD{\Agent{A}}{\Agent{C}} \nAnd \pfc{\Agent{C}}{\Agent{B}}}$. This means that for a given \soca, if there exists a sequence of agents, starting with a source agent~$\Agent{A}$ and ending on a sink agent~$\Agent{B}$, that have the \PFCD either via external stimuli or via shared environments, then~$\Agent{A}$ has the \PFC with~$\Agent{B}$. 

A useful result showing the effects of modifying the behaviour of an agent in the sequence of a potential communication path between two agents is given in Proposition~\ref{prop:general_PFC}. Recall that we say that a stimulus generated by an agent~$\Agent{A}$ \emph{influences} an agent~$\Agent{B}$ if the behaviour of~$\Agent{B}$ changes as a result of the response to the stimulus (\ie~$\biglnotation{\exists}{s}{s \in \STIMset}{\lAct{b}{\lOut{a}{s}} \neq b}$).

\begin{proposition}
\label{prop:general_PFC}
	Let~$\pfc{\Agent{A}}{\Agent{B}}$ such that~$\biglnotation{\exists}{\Agent{C}}{\Agent{C} \in \C}{\pfcD{\Agent{A}}{\Agent{C}} \nAnd \pfc{\Agent{C}}{\Agent{B}}}$ where~$\agent{A}{a}$,~$\agent{B}{b}$, and~$\agent{C}{c}$. Let~$\depOp$ be the given dependence relation. Suppose~$\Agent{C}$ is replaced by another agent~$\agent{C'}{c'}$. Then,
	\begin{enumerate}[(i)]
		\item \label{prop:PFC_seq} 
			If~$c' = (c \CKAseq d)$, then~$\pfc{\Agent{A}}{\Agent{B}}$ only if~$\bigP{\dep{(c \CKAseq d)}{a} \nAnd \dep{b}{(c \CKAseq d)}} \Ors \biglnotation{\exists}{t}{t \in \STIMset}{\lAct{b}{\lOut{(c \CKAseq d)}{t}} \neq b}$.
		\item \label{prop:PFC_plus} 
			If~$c' = (c + d)$, then~$\pfc{\Agent{A}}{\Agent{B}}$ only if~$\biglnotation{\forall}{t}{t \in \STIMbasic}{\Not(\lAct{d}{t} \CKAle c + d)}$.
		\item \label{prop:PFC_iterSeq} 
			If~$c' = \CKAiterSeq{c}$, then~$\pfc{\Agent{A}}{\Agent{B}}$.
		\item \label{prop:PFC_inactive_idle} 
			If~$c' = 0$ or~$c' = 1$ and the \emph{\CCKAabbrv} is without reactivation, then~$\notpfc{\Agent{A}}{\Agent{B}}$.
		\item \label{prop:PFC_orbit} 
			If~$c' \in \orbS{c}$, then~$\pfc{\Agent{A}}{\Agent{B}}$.
		\item \label{prop:PFC_fixed} 
			If~$c'$ is a fixed point behaviour, then~$\pfc{\Agent{A}}{\Agent{B}}$ only if~$\dep{c'}{a} \nAnd \dep{b}{c'}$.
	\end{enumerate}
	
	\begin{proof}
		Each of the proofs involve the applications of definitions of~$\pfcD{}{}$,~$\STIMcommD{}$, and~$\ENVcommD{}{}$ as well as the basic axioms of \CCKAabbrv. The detailed proofs can be found in Appendix~\ref{sub:detailed_proof_of_proposition_general_PFC}.
	\end{proof}
\end{proposition}

Proposition~\ref{prop:general_PFC} identifies the conditions constraining the modifications allowable to the behaviour of an agent in a potential communication path in order to maintain the \PFC between two agents. Specifically, Identity~(\ref{prop:PFC_seq}) shows how the sequential composition of an additional behaviour with the existing agent will not affect the \PFC provided the composed behaviour preserves the dependency relation or has the ability to influence the behaviour of the next agent in the path. Assuming that each agent behaviour takes some amount of time, this is useful since we can construct behaviours that satisfy this constraint to introduce delay into the potential communication path in order to disturb a covert timing channel without the need to fully eliminate the communication. However, in general, we cannot say anything about the behaviour~$d$ alone as a consequence of Definition~\ref{def:C2KA}(\ref{def:cascading_output_axiom}). The stimuli that are generated by~$d$ are dependent on the stimuli generated by~$c$ and the effects of the stimuli cascaded from~$c$ to~$d$ cannot be determined since~$\Agent{C'}$ is viewed as a black-box. Identity~(\ref{prop:PFC_plus}) is an extension of Propositions~\ref{prop:stim_comm_plus} and~\ref{prop:env_comm_plus} to general \PFC. In general, provided that the introduction of non-determinism does not result in a fixed point behaviour, the \PFC is maintained with the addition of non-determinism. Identity~(\ref{prop:PFC_iterSeq}) follows from Identities~(\ref{prop:PFC_seq}) and~(\ref{prop:PFC_plus}) and shows that the sequential iteration of an agent behaviour does not affect the \PFC. Identity~(\ref{prop:PFC_inactive_idle}) states that if we replace an agent in a communication path with an inactive agent or an idle agent when we have a \CCKAabbrv without reactivation, then there is no longer a \PFC. This can be useful in terms of eliminating the \PFC among agents since it shows how we may modify the behaviour of some agents in order to eliminate the \PFC and potentially thwart any attempts for establishing covert communication channels. However, it is noted that this is not a suitable solution in all cases since modifying agent behaviours in such a way can inadvertently modify the overall system behaviour and thereby undesirably render the system useless. Identity~(\ref{prop:PFC_orbit}) states that replacing an agent in a given communication path with another agent in the same strong orbit will not affect the \PFC. This is because agents in the same strong orbit always have the \PFC via external stimuli with one another. Identity~(\ref{prop:PFC_fixed}) states that the \PFC is maintained when replacing an agent in a given communication path with another agent that has a fixed point behaviour only if the dependency relation is preserved. Proposition~\ref{prop:fixed_point} showed that an agent with a fixed point behaviour does not have the \PFC via external stimuli unless it is the source of a potential communication path. So, if an agent with a fixed point behaviour is not the source of the potential communication path, then it may only have the \PFC via shared environments. Finally, it should be noted that if we restrict the behaviour of an agent in a potential communication path to a particular sub-behaviour, then the \PFC is only preserved if the sub-behaviour maintains the communicating behaviour of the original agent. 


\vspace{-0.175in}
\section{Discussion and Related Work}
\label{sec:discussion}
\vspace{-0.1in}
Given a \soca, it is difficult to fully prevent the possibility of covert communication from taking place since it is often undesirable to completely eliminate the communication among agents. An integral part of safeguarding \socas from covert channel communication is having the ability to identify when a covert channel may exist in a given system which involves determining if and when two agents have a \PFC. While much of the existing work in attempting to mitigate covert channels has been based on information theoretic approaches~(\eg~\cite{Gianvecchio2007aa,Giles2002aa,Giles1999aa,Helouet2010aa,Lowe2002aa,Martin2006aa,Millen1987aa,Millen1989aa,Moskowitz1994aa}), the proposed formulation looks to the issue of mitigating covert channels from a different perspective. Although, it is difficult to completely eliminate covert channels from modern computer systems, the proposed formalisation provides a means for analysing a \soca in order to devise mechanisms for strengthening the design of such systems in order to make them more robust against covert channels. It also builds the foundation for the ability to identify parts of a system where it would be most beneficial to observe or disrupt the communication among particular system agents. For example, once we have identified a sequence of agents that have the \PFC, in order to detect confidential information leakage via protocol-based covert channels, we can install monitors that are configured to identify patterns of communication on the communication channels available to the agents in the potential communication path using techniques similar to that presented in~\cite{Jaskolka2011ac}. Similarly, in order to mitigate the use of covert timing channels, we can employ mechanisms that de-couple or deteriorate any sort of timing information associated with the communication channels available to the agents in the potential communication path by injecting random delays similar to the NRL Pump~\cite{Kang1993aa}. 

In the literature, we find existing works that have attempted to articulate and verify potential for communication conditions for covert channels. However, some of them are indirect or informal and require reasoning about potential scenarios in which the conditions might be satisfied (\eg~\cite{Shieh1999aa}). Furthermore, those works which do provide some level of formalism, focus primarily on the potential for communication via shared environments through various information flow analyses based on finite state machine models, information theory, and probability theory (\eg~\cite{Gray1991aa,Johnson2010aa,Millen1989aa,Wang2005aa}). Perhaps one of the most popular mechanisms for determining the \PFC for identifying the existence of covert channels is the Shared Resource Matrix technique~\cite{Kemmerer1983aa}. It involves a careful analysis of the ways in which shared resources are used in a system to determine whether it is possible for a particular resource to covertly transfer information from one agent to another with respect to a set of minimum criteria. Similarly, Covert Flow Trees (\eg~\cite{Kemmerer1991aa}) attempt to identify information flows supporting either the direct or indirect ability of an agent to detect when an attribute of a shared resource has been modified. The Shared Resource Matrix technique and Covert Flow Trees can be used in our formulation to concretely build the dependence relation discussed in paragraph~$3$ of Section~\ref{sub:potential_for_communication_via_shared_environments}.

While existing works focus on studying the potential for communication via shared environments, the proposed formulation of the \PFC condition for covert channel existence is based on the mathematical foundation of \CCKAabbrv and thereby also considers the potential for communication via external stimuli. If we were to consider the use of \CKAabbrv alone for the formulation of the \PFC condition, we can only use the dependencies between shared events to define and verify any sort of \PFC. The proposed formulation provides a more complete representation of the potential means for communication among system agents that encompasses what can be done using \CKAabbrv alone as well as other existing information flow techniques.  

\vspace{-0.15in}
\section{Conclusion and Future Work} 
\label{sec:conclusion_and_future_work}
\vspace{-0.1in}
In this paper, we presented a formulation of the \PFC condition for covert channel existence. The proposed formulation is based on the mathematical framework of \CCKA (\CCKAabbrv). It allows for the consideration of the \PFC from the perspective of shared environments as well as the perspective of external stimuli. To the best of our knowledge, there does not exist a formulation of the \PFC in \socas that considers the \PFC via both external stimuli and shared environments. The proposed formulation and its mathematical background help to analyse \socas in order to devise mechanisms for strengthening such systems against covert channels.

In future work, we aim to support the automated verification of the \PFC condition for covert channel existence. We are developing tool support to aid in the specification and verification of the \PFC condition for \socas. We are also investigating the adaptation of description logic~\cite{Baader2003aa} to develop a formulation of the \COC condition for covert channel existence~\cite{Jaskolka2012aa} in \socas. Then, we aim to propose guidelines for designing \socas that are resilient to covert channels.

\vspace{-0.2in}
\bibliographystyle{eptcs} 
\bibliography{Bibliography/GandALF2014}

\vspace{-0.2in}
\appendix
\section{Detailed Proofs of Propositions}
\label{sec:detailed_proofs}
\begingroup
\fontsize{11pt}{8pt}\selectfont
\vspace{-0.1in}
\paragraph{Detailed Proof of Proposition~\ref{prop:universally_influential_connected}:}
\label{sub:detailed_proof_of_proposition_universally_influential_connected}

Assume~$\C$ is a stimuli-disconnected system and let~$\Agent{C} \in \C$ be universally influential. Also, assume that there exists a partition of~$\C$,~$X_1$ and~$X_2$, such that~$\Agent{C} \in X_2$. We prove by contradiction.		

\Beginproof
	\pred{$\C$ is stimuli-disconnected~$\nAnd \Agent{A}$ is universally influential}
	$\mIff$ \com{Definition of Stimuli-Disconnected}
	\pred{$\lnotation{\forall}{\Agent{A},\Agent{B}}{\Agent{A} \in X_1 \nAnd \Agent{B} \in X_2}{\notSTIMcomm{\Agent{A}}{\Agent{B}} \nAnd \notSTIMcomm{\Agent{B}}{\Agent{A}}} \nAnd \Agent{C}$ is universally influential}
	$\mImp$ \com{Instantiation:~$\Agent{B} = \Agent{C}$}
	\pred{$\lnotation{\forall}{\Agent{A}}{\Agent{A} \in X_1}{\notSTIMcomm{\Agent{A}}{\Agent{C}} \nAnd \notSTIMcomm{\Agent{C}}{\Agent{A}}} \nAnd \Agent{C}$ is universally influential}
	$\mImp$ \com{Definition of Universally Influential}
	\pred{$\lnotation{\forall}{\Agent{A}}{\Agent{A} \in X_1}{\notSTIMcomm{\Agent{A}}{\Agent{C}} \nAnd \false}$}
	$\mIff$ \com{Zero of~$\nAnd$ \hsep~$\forall$-False Body}
	\pred{$\false$}
\Endproof


\vspace{-0.35in}
\paragraph{Detailed Proof of Proposition~\ref{prop:stim_comm_plus}:}
\label{sub:detailed_proof_of_proposition_stim_comm_plus}

Let~$\agent{A}{a}$,~$\agent{B}{b}$, and~$\agent{C}{c}$ be agents in~$\C$.
\begin{enumerate}[(i)]
	\item  If~$\STIMcommD{\Agent{B}}{\Agent{C}}$ then~$\STIMcommD{(\Agent{A} + \Agent{B})}{\Agent{C}}$.
	\Beginproof
		\pred{$\STIMcommD{(\Agent{A} + \Agent{B})}{\Agent{C}}$}
		$\mIff$ \com{Definition of~$\STIMcommD{}{}$}
		\pred{$\biglnotation{\exists}{s,t}{s,t \in \STIMbasic \nAnd t \STIMle \lOut{a + b}{s}}{\lAct{c}{t} \neq c}$}
		$\mIff$ \com{Distributivity of~$\outOp$ over~$+$}
		\pred{$\biglnotation{\exists}{s,t}{s,t \in \STIMbasic \nAnd t \STIMle \lOut{a}{s} \STIMplus \lOut{b}{s}}{\lAct{c}{t} \neq c}$}
		$\mImpl$ \com{Definition of~$\STIMle$ \hsep Isotony of~$=$}
		\pred{$\biglnotation{\exists}{s,t}{s,t \in \STIMbasic \nAnd t \STIMle \lOut{b}{s}}{\lAct{c}{t} \neq c}$}
		$\mImpl$ \com{Hypothesis:~$\STIMcommD{\Agent{B}}{\Agent{C}}$}
		\pred{$\true$}
	\Endproof
\vspace{-0.15in}
	\item If~$\STIMcommD{\Agent{A}}{\Agent{B}}$ then~$\STIMcommD{\Agent{A}}{(\Agent{B} + \Agent{C})}$ only if~$\lnotation{\forall}{t}{t \in \STIMbasic}{\Not(\lAct{c}{t} \CKAle b + c)}$.
	\Beginproof
		\pred{$\STIMcommD{\Agent{A}}{(\Agent{B} + \Agent{C})}$}
		$\mIff$ \com{Definition of~$\STIMcommD{}{}$}
		\pred{$\biglnotation{\exists}{s,t}{s,t \in \STIMbasic \nAnd t \STIMle \lOut{a}{s}}{\lAct{(b + c)}{t} \neq b + c}$}
		$\mIff$ \com{Distributivity of~$\actOp$ over~$+$}
		\pred{$\biglnotation{\exists}{s,t}{s,t \in \STIMbasic \nAnd t \STIMle \lOut{a}{s}}{\lAct{b}{t} + \lAct{c}{t} \neq b + c}$}
		$\mImpl$ \com{Weakening}
		\pred{$\biglnotation{\exists}{s,t}{s,t \in \STIMbasic \nAnd t \STIMle \lOut{a}{s}}{\Not(\lAct{c}{t} + \lAct{b}{t} \CKAle b + c)}$}
		$\mIff$ \com{Definition of~$\CKAle$ \hsep Idempotence of~$+$}
		\pred{$\biglnotation{\exists}{s,t}{s,t \in \STIMbasic \nAnd t \STIMle \lOut{a}{s}}{\Not(\lAct{c}{t} + b + c + \lAct{b}{t} = b + b + c)}$}
		$\mIff$ \com{Isotony of~$=$}
		\pred{$\biglnotation{\exists}{s,t}{s,t \in \STIMbasic \nAnd t \STIMle \lOut{a}{s}}{\Not(\lAct{c}{t} + b + c  = b + c \Ors \lAct{b}{t} = b)}$}
		$\mIff$ \com{De Morgan}
		\pred{$\biglnotation{\exists}{s,t}{s,t \in \STIMbasic \nAnd t \STIMle \lOut{a}{s}}{\lAct{c}{t} + b + c \neq b + c \nAnd \lAct{b}{t} \neq b}$}
		$\mImpl$ \com{Hypothesis:~$\STIMcommD{\Agent{A}}{\Agent{B}} \!\!\!\mImp\!\!\! \lAct{b}{t} \neq b$ \!\!\!\!\!\!\hsep\!\!\!\!\!\! Hypothesis:~$\lnotation{\forall}{t\!\!}{\!\!t \in \STIMbasic\!\!}{\!\!\Not(\lAct{c}{t} \CKAle b + c)}$\!\!}
		\pred{$\biglnotation{\exists}{s,t}{s,t \in \STIMbasic \nAnd t \STIMle \lOut{a}{s}}{\true}$}
		$\mIff$ \com{$\exists$-True Body}
		\pred{$\true$}
	\Endproof
\end{enumerate}


\vspace{-0.35in}
\paragraph{Detailed Proof of Proposition~\ref{prop:general_PFC}:}
\label{sub:detailed_proof_of_proposition_general_PFC}

Let~$\pfc{\Agent{A}}{\Agent{B}}$ such that~$\biglnotation{\exists}{\Agent{C}\!\!}{\!\!\Agent{C} \in \C}{\pfcD{\Agent{A}}{\Agent{C}} \!\!\nAnd\!\! \pfc{\Agent{C}}{\Agent{B}}}$. 
For simplicity, we assume that~$\pfc{\Agent{A}}{\Agent{B}}$ via~$\pfcD{\Agent{A}}{\Agent{C'}} \!\!\nAnd\!\! \pfcD{\Agent{C'}}{\Agent{B}}$ unless stated otherwise.

\begin{enumerate}[(i)]
	\item~$\agent{C'}{c \CKAseq d}$
\vspace{-0.05in}	
	\Beginproof
		\pred{$\pfcD{\Agent{A}}{\Agent{C'}} \nAnd \pfc{\Agent{C'}}{\Agent{B}}$}
		$\mIff$ \com{Definition of~$\pfc{}{}$}
		\pred{$\pfcD{\Agent{A}}{\Agent{C'}} \nAnd \bigP{\STIMcomm{\Agent{C'}}{\Agent{B}} \Ors \ENVcommD{\Agent{C'}}{\Agent{B}}}$}
		$\mIff$ \com{Definition of~$\pfcD{}{}$}
		\pred{$\bigP{\STIMcommD{\Agent{A}}{\Agent{C'}} \Ors \ENVcommD{\Agent{A}}{\Agent{C'}}} \nAnd \bigP{\STIMcomm{\Agent{C'}}{\Agent{B}} \Ors \ENVcommD{\Agent{C'}}{\Agent{B}}}$}
		$\mIff$ \com{Definition of~$\STIMcommD{}{}$ \hsep Definition of~$\ENVcommD{}{}$}
		\pred{$\bigP{\biglnotation{\exists}{s,t\!\!\!}{\!\!\!s,t \in \STIMbasic \!\!\nAnd\!\! t \STIMle \lOut{a}{s}\!\!}{\!\!\lAct{(c \CKAseq d)}{t} \!\!\neq\!\! c \CKAseq d} \!\!\Ors\!\! \dep{(c \CKAseq d)}{a}} \!\!\nAnd\!\! \bigP{\STIMcomm{\Agent{C'}}{\Agent{B}} \!\!\Ors\!\! \dep{b}{(c \CKAseq d)}}$}	
		$\mIff$ \com{Definition~\ref{def:C2KA}(\ref{def:cascading_axiom})}
		\pred{$\bigP{\biglnotation{\exists}{s,t}{s,t \in \STIMbasic \nAnd t \STIMle \lOut{a}{s}}{(\lAct{c}{t}) \CKAseq \bigP{\lAct{d}{\lOut{c}{t}}} \neq c \CKAseq d} \Ors \dep{(c \CKAseq d)}{a}} \nAnd \\ \bigP{\STIMcomm{\Agent{C'}}{\Agent{B}} \Ors \dep{b}{(c \CKAseq d)}}$}		
		$\mIff$	\com{Definition of $\neq$}
		\pred{$\bigP{\biglnotation{\exists}{s,t}{s,t \in \STIMbasic \nAnd t \STIMle \lOut{a}{s}}{\Not\bigP{(\lAct{c}{t}) \CKAseq \bigP{\lAct{d}{\lOut{c}{t}}} = c \CKAseq d}} \Ors \dep{(c \CKAseq d)}{a}} \nAnd \\ \bigP{\STIMcomm{\Agent{C'}}{\Agent{B}} \Ors \dep{b}{(c \CKAseq d)}}$}	
		$\mImpl$ \com{Isotony of $=$}
		\pred{$\bigP{\biglnotation{\exists}{s,t}{s,t \in \STIMbasic \nAnd t \STIMle \lOut{a}{s}}{\Not\bigP{\lAct{c}{t} = c \nAnd \bigP{\lAct{d}{\lOut{c}{t}}} = d}} \Ors \dep{(c \CKAseq d)}{a}} \nAnd \\ \bigP{\STIMcomm{\Agent{C'}}{\Agent{B}} \Ors \dep{b}{(c \CKAseq d)}}$}	
		$\mIff$ \com{De Morgan}
		\pred{$\bigP{\biglnotation{\exists}{s,t}{s,t \in \STIMbasic \nAnd t \STIMle \lOut{a}{s}}{\lAct{c}{t} \neq c \Ors \bigP{\lAct{d}{\lOut{c}{t}}} \neq d} \Ors \dep{(c \CKAseq d)}{a}} \nAnd \\ \bigP{\STIMcomm{\Agent{C'}}{\Agent{B}} \Ors \dep{b}{(c \CKAseq d)}}$}	
		$\mImpl$ \com{Hypothesis:~$\bigP{\dep{(c \CKAseq d)}{a} \!\!\!\nAnd\!\!\! \dep{b}{(c \CKAseq d)}} \!\!\!\Ors\!\!\! \biglnotation{\exists}{t}{t \in \STIMset}{\lAct{b}{\lOut{(c \CKAseq d)}{t}} \!\neq\! b}$ \hsep $\pfcD{\Agent{A}}{\Agent{C}} \!\!\mImp\!\! \lnotation{\exists}{t\!\!}{\!\!\!t \in \STIMbasic}{\lAct{c}{t} \neq c}$}
		\pred{$\true$}
	\Endproof
\vspace{-0.15in}	
	\item~$\agent{C'}{c + d}$
	\Beginproof
		\pred{$\pfcD{\Agent{A}}{\Agent{C'}} \nAnd \pfcD{\Agent{C'}}{\Agent{B}}$}
		$\mIff$ \com{Definition of~$\pfcD{}{}$}
		\pred{$\bigP{\STIMcommD{\Agent{A}}{\Agent{C'}} \Ors \ENVcommD{\Agent{A}}{\Agent{C'}}} \nAnd \bigP{\STIMcommD{\Agent{C'}}{\Agent{B}} \Ors \ENVcommD{\Agent{C'}}{\Agent{B}}}$}
		$\mIff$ \com{Definition of~$\STIMcommD{}{}$ \hsep Definition of~$\ENVcommD{}{}$}
		\pred{$\bigP{\biglnotation{\exists}{s,t}{s,t \in \STIMbasic \nAnd t \STIMle \lOut{a}{s}}{\lAct{(c + d)}{t} \neq c + d} \Ors \dep{(c + d)}{a}} \nAnd \\ \bigP{\biglnotation{\exists}{s,t}{s,t \in \STIMbasic \nAnd t \STIMle \lOut{(c + d)}{s}}{\lAct{b}{t} \neq b} \Ors \dep{b}{(c + d)}}$}
		$\mImpl$ \com{Hypothesis:~$\pfcD{\Agent{A}}{\Agent{C}} \nAnd \pfcD{\Agent{C}}{\Agent{B}}$ \hsep Hypothesis:~$\lnotation{\forall}{t}{t \in \STIMbasic}{\Not(\lAct{d}{t} \CKAle c + d)}$ \hsep Proposition~\ref{prop:stim_comm_plus} \hsep Proposition~\ref{prop:env_comm_plus}}
		\pred{$\true$}
	\Endproof
\vspace{-0.15in}	
	\item~$\agent{C'}{\CKAiterSeq{c}}$
	\Beginproof
		\pred{$\pfcD{\Agent{A}}{\Agent{C'}} \nAnd \pfcD{\Agent{C'}}{\Agent{B}}$}
		$\mIff$ \com{Definition of~$\pfcD{}{}$}
		\pred{$\bigP{\STIMcommD{\Agent{A}}{\Agent{C'}} \Ors \ENVcommD{\Agent{A}}{\Agent{C'}}} \nAnd \bigP{\STIMcommD{\Agent{C'}}{\Agent{B}} \Ors \ENVcommD{\Agent{C'}}{\Agent{B}}}$}
		$\mIff$ \com{Definition of~$\STIMcommD{}{}$ \hsep Definition of~$\ENVcommD{}{}$}
		\pred{$\bigP{\biglnotation{\exists}{s,t}{s,t \in \STIMbasic \nAnd t \STIMle \lOut{a}{s}}{\lAct{\CKAiterSeq{c}}{t} \neq \CKAiterSeq{c}} \Ors \dep{\CKAiterSeq{c}}{a}} \nAnd \\ \bigP{\biglnotation{\exists}{s,t}{s,t \in \STIMbasic \nAnd t \STIMle \lOut{\CKAiterSeq{c}}{s}}{\lAct{b}{t} \neq b} \Ors \dep{b}{\CKAiterSeq{c}}}$}
		$\mImpl$ \com{Definition of~$\CKAiterSeq{}$ \hsep Proposition~\ref{prop:general_PFC}(\ref{prop:PFC_plus})}
		\pred{$\true$}
	\Endproof
\vspace{-0.15in}	
	\item~$\agent{C'}{0}$
	\Beginproof
		\pred{$\pfcD{\Agent{A}}{\Agent{C'}} \nAnd \pfcD{\Agent{C'}}{\Agent{B}}$}
		$\mIff$ \com{Definition of~$\pfcD{}{}$}
		\pred{$\bigP{\STIMcommD{\Agent{A}}{\Agent{C'}} \Ors \ENVcommD{\Agent{A}}{\Agent{C'}}} \nAnd \bigP{\STIMcommD{\Agent{C'}}{\Agent{B}} \Ors \ENVcommD{\Agent{C'}}{\Agent{B}}}$}
		$\mIff$ \com{$0$ is a fixed point behaviour \!\!\!\!\!\!\hsep\!\!\!\!\!\! Proposition~\ref{prop:fixed_point} \!\!\!\!\!\hsep\!\!\!\!\! $\indep{0}{a}$}
		\pred{$\bigP{\false \Ors \false} \nAnd \bigP{\STIMcommD{\Agent{C'}}{\Agent{B}}{} \Ors \ENVcommD{\Agent{C'}}{\Agent{B}}{}}$}
		$\mIff$ \com{Idempotence of~$\Ors$ \hsep Zero of~$\nAnd$}
		\pred{$\false$}
	\Endproof
\vspace{-0.15in}	
	The proof is similar when~$\agent{C'}{1}$ and the \CCKAabbrv is without reactivation (\ie $\lnotation{\forall}{s}{s \in \STIMset \STdiff \set{\Dstim}}{\lAct{s}{1} = 1}$).~\newline 
	\item~$\agent{C'}{c'}$ such that~$c' \in \orbS{c}$
	\Beginproof
		\pred{$\pfcD{\Agent{A}}{\Agent{C'}} \nAnd \pfcD{\Agent{C'}}{\Agent{B}}$}
		$\mIff$ \com{Definition of~$\pfcD{}{}$}
		\pred{$\bigP{\STIMcommD{\Agent{A}}{\Agent{C'}} \Ors \ENVcommD{\Agent{A}}{\Agent{C'}}} \nAnd \bigP{\STIMcommD{\Agent{C'}}{\Agent{B}} \Ors \ENVcommD{\Agent{C'}}{\Agent{B}}}$}
		$\mImpl$ \com{Hypothesis:~$\pfcD{\Agent{A}}{\Agent{C}} \nAnd \pfcD{\Agent{C}}{\Agent{B}}$ \hsep Hypothesis:~$c' \in \orbS{c} \mImp \\ \lnotation{\exists}{s,t}{s,t \in \STIMset}{\lAct{c}{s} = c' \nAnd \lAct{c'}{t} = c} \mImp \STIMcomm{\Agent{C}}{\Agent{C'}} \nAnd \STIMcomm{\Agent{C'}}{\Agent{C}}$}
		\pred{$\true$}	
	\Endproof
\vspace{-0.15in}
	\item~$\agent{C'}{c'}$ such that~$c'$ is a fixed point behaviour
	\Beginproof
		\pred{$\pfcD{\Agent{A}}{\Agent{C'}} \nAnd \pfcD{\Agent{C'}}{\Agent{B}}$}
		$\mIff$ \com{Definition of~$\pfcD{}{}$}
		\pred{$\bigP{\STIMcommD{\Agent{A}}{\Agent{C'}} \Ors \ENVcommD{\Agent{A}}{\Agent{C'}}} \nAnd \bigP{\STIMcommD{\Agent{C'}}{\Agent{B}} \Ors \ENVcommD{\Agent{C'}}{\Agent{B}}}$}
		$\mIff$ \com{Definition of~$\ENVcommD{}{}$}
		\pred{$\bigP{\STIMcommD{\Agent{A}}{\Agent{C'}} \Ors \dep{c'}{a}} \nAnd \bigP{\STIMcommD{\Agent{C'}}{\Agent{B}} \Ors \dep{b}{c'}}$}
		$\mImpl$ \com{Hypothesis:~$c'$ is a fixed point behaviour \!\!\!\!\!\!\!\hsep\!\!\!\!\!\!\! Proposition~\ref{prop:fixed_point}}
		\pred{$\bigP{\false \Ors \dep{c'}{a}} \nAnd \bigP{\STIMcommD{\Agent{C'}}{\Agent{B}} \Ors \dep{b}{c'}}$}
		$\mIff$ \com{Identity of~$\Ors$}
		\pred{$\dep{c'}{a} \nAnd \bigP{\STIMcommD{\Agent{C'}}{\Agent{B}} \Ors \dep{b}{c'}}$}
		$\mImpl$ \com{Hypothesis:~$\dep{c'}{a} \nAnd \dep{b}{c'}$} 
		\pred{$\true \nAnd \bigP{\STIMcommD{\Agent{C'}}{\Agent{B}} \Ors \true}$}
		$\mIff$ \com{Zero of~$\Ors$ \hsep Idempotence of~$\nAnd$}
		\pred{$\true$}
	\Endproof
\end{enumerate}

\endgroup

\end{document}